\documentclass[useAMS,usenatbib]{mn2e}
\usepackage{epsfig}

\usepackage{float}

\usepackage{lipsum}
\usepackage{longtable,amsmath}

\title[The Main Sequence of three RSGCs]{The Main Sequence of three Red Supergiant
Clusters}

\author[Froebrich \& Scholz]{Dirk Froebrich$^{1}$\thanks{E-mail:
df@star.kent.ac.uk}, Alexander Scholz$^{2,3}$\\$^{1}$Centre for Astrophysics and
Planetary Science, University of Kent, Canterbury, CT2 7NH, United
Kingdom\\$^{2}$Dublin Institute for Advanced Studies, 5 Merrion Square, Dublin
2, Ireland\\$^{3}$School of Physics and Astronomy, University of St. Andrews,
North Haugh, St. Andrews, KY16 9SS, United Kingdom}

\begin{document}

\date{Received today / Accepted tomorrow}

\pagerange{\pageref{firstpage}--\pageref{lastpage}} \pubyear{2012}

\maketitle

\label{firstpage}

\begin{abstract}

Massive clusters in our Galaxy are an ideal testbed to investigate the
properties and evolution of high mass stars. They provide statistically
significant samples of massive stars of uniform ages. To accurately determine
the intrinsic physical properties of these stars we need to establish the
distances, ages and reddening of the clusters. One avenue to achieve this is the
identification and characterisation of the main sequence members of red
supergiant rich clusters. 

Here we utilise publicly available data from the UKIDSS galactic plane survey.
We show that point spread function photometry in conjunction with standard
photometric decontamination techniques allows us to identify the most likely
main sequence members in the 10-20\,Myr old clusters RSGC\,1, 2, and 3. We
confirm the previous detection of the main sequence in RSGC\,2 and provide the
first main sequence detection in RSGC\,1 and RSGC\,3. There are in excess of 100
stars with more than 8\,$M_\odot$ identified in each cluster. These main
sequence members are concentrated towards the spectroscopically confirmed red
supergiant stars. We utilise the $J$$-$$K$ colours of the bright main sequence
stars to determine the $K$-band extinction towards the clusters. The
differential reddening is three times as large in the youngest cluster RSGC\,1
compared to the two older clusters RSGC\,2 and RSGC\,3. Spectroscopic follow up
of the cluster main sequence stars should lead to more precise distance and age
estimates for these clusters as well as the determination of the stellar mass
function in these high mass environments.

\end{abstract}


\section{Introduction}\label{intro}
 
Massive stars are most commonly formed in clusters and associations
\citep{2012MNRAS.424.3037G}, even if there are potential exceptions (e.g.
\citet{2013ApJ...768...66O}). These building blocks are the sole observational
characteristic of star formation that is observable in more distant galaxies. It
is thus of importance to investigate the details of such massive clusters
locally in our own Galaxy. 

The identification of local examples of clusters and associations of massive
stars is complicated by a number of facts. Such objects are generally rare in
the Milky Way and are thus typically at large distances. The Sun's position near
the Galactic Plane, hence makes it difficult to identify and investigate these
objects due to the large amounts of extinction along the line of sight.
Furthermore, most massive star formation is projected towards the general
direction of the Galactic Centre, which causes additional difficulties due to
crowding.

Thus, the identification and characterisation of massive star clusters in our
own Galaxy has only recently gained considerable momentum due to advances in
infrared astronomy that allow us to probe these distant and obscured objects.
Many of them have been known for a considerable time, but have not been
recognised as massive star clusters. For example Westerlund\,1 was discovered by
\citet{1961PASP...73...51W} but only recognised as young, very massive cluster
more than 30\,yrs later \citep{1998MNRAS.299L..43C, 1998A&AS..127..423P}.
Similarly, the cluster Stephenson\,2 \citep{1990AJ.....99.1867S} had been known
for many years until its true nature as Red Supergiant Cluster\,2 (RSGC\,2) was
uncovered \citep{2002A&A...390..931O, 2007ApJ...671..781D}. Further, more
systematic searches to uncover the population of the most massive clustes in the
Galaxy are ongoing, e.g. \citet{2013ApJ...766..135R}. 

Clusters at the top end of the mass distribution in our Galaxy (above
$10^4$\,M$\odot$) have, depending on their age, a sizable number of evolved
massive stars. These are either Wolf-Rayet stars, or blue/yellow/red
supergiants. There are now a number of galactic clusters with a large population
of Red Supergiant (RSG) stars known, many of which are in a region confined to
where the Scutum spiral arm meets the near side of the Galactic Bar. These are
e.g. RSGC\,1 to RSGC\,5 \citep{2006ApJ...643.1166F, 2007ApJ...671..781D,
2009A&A...498..109C, 2009AJ....137.4824A, 2010A&A...513A..74N,
2011A&A...528A..59N}. All these clusters are about 6\,kpc from the Sun, highly
extincted, between 7\,Myr and 20\,Myr old and have masses in excess of
10.000\,$M_\odot$. These clusters are thus an ideal testbed to study the
formation and evolution of massive stars as well as their influence on the
environment in great detail. However, it is important in the context of
understanding the formation and evolution of  massive clusters to investigate
the distribution and properties of the lower mass stars as well as the bright
supergiants and Wolf Rayet stars. In particular since the intrinsic properties
of lower mass stars are much better understood, this should enable us to
determine the distances, ages and reddening of these clusters more accurately.

The main sequence (MS) stars of these clusters have so far, however, not been
investigated (though the slightly less massive and less reddened cluster
NGC\,7419 has recently been studied by \citet{2013A&A...552A..92M}). Given that
the RSG cluster members have an apparent brightness of about $K$\,=\,6\,mag, one
would expect to detect the MS in deep near-infrared (NIR) images, in particular
the Galactic Plane Survey (GPS, \citet{2008MNRAS.391..136L}) data from the UK
Deep Sky Survey (UKIDSS, \citet{2007MNRAS.379.1599L})  seems to be ideal to
detect the fainter members of the RSG clusters mentioned above. But even data
from the 2--Micron All Sky Survey (2MASS, \citet{2006AJ....131.1163S})  should
be deep enough to detect the brightest main sequence stars in these clusters. 

Only for the cluster RSGC\,2 has the main sequence been detected so far
\citep{2013IJAA...03..161F}. The author uses 2MASS and GPS data to identify the
top of the MS at colours of about $J$$-$$K$\,=\,1.5\,mag and at a brightness of
slightly fainter than $K$\,=\,10\,mag. The author also investigates the data for
RSGC\,1 and RSGC\,3 with the same methods but is unable to detect the main
sequence for these clusters. It is speculated that this is caused by problems
with accurate aperture photometry in the vicinity of the bright RSG stars in
those objects. The main sequence in RSGC\,2 is the easiest to detect due to the
spatial extent of the cluster, which is the largest amongst the RSG clusters.
Here we hence try to identify the main sequences of other known and candidate
RSG clusters by means of point spread function (PSF) photometry in the JHK
images of the UKIDSS GPS data. This will include some of the new RSG cluster
candidates (F\,3 and F\,4) identified by \citep{2013IJAA...03..161F} based on
colour selected star density maps from 2MASS.

This paper is structured as follows. In Sect.\,\ref{data} we briefly outline the
data and analysis methods employed. We then discuss our results in
Sect.\,\ref{results} with focus on the properties of the main sequences detected
in RSGC\,1, 2, 3.

 %
 %
 %
 %
 %
 %
 %
 %
 %

\begin{table*}
\centering

\caption{\label{calibration}  Here we list the clusters and candidates
investigated and their central coordinates. We also indicate the total number of
stars in the  10\arcmin\,$\times$\,10\arcmin\ field around each cluster position
in our input catalogue and the number of stars with good PSF photometry in all
three bands that are brighter than the completeness limit in each filter (see
text for details). We also list the $rms$ scatter of the photometric calibration
of our PSF photometry into the UKIDSS system. The $min$ and $max$ columns give
the range of magnitudes for which the calibration was performed. The $\sigma$
values denote the $rms$ scatter for each field, filter and specified magnitude
range.} 

\begin{tabular}{c|cc|cc|ccc|ccc|ccc}
Field & RA & DEC & \multicolumn{2}{c}{Number of stars} & $J_{min}$ &  $J_{max}$ & $\sigma_J$  & $H_{min}$ &  $H_{max}$ &
$\sigma_H$ & $K_{min}$ &  $K_{max}$ & $\sigma_K$ \\ 
 & \multicolumn{2}{c}{[deg] (J2000)} & Input & Output & [mag] & [mag] & [mag] & [mag] & [mag] & [mag] & [mag] & [mag] & [mag] \\ \hline 
\noalign{\smallskip}
RSGC\,1  & 279.488 & $-$6.880 & 31562 & 11976 & 14.0 & 17.0 & 0.024 & 13.0 & 16.0 & 0.034 & 12.0 & 15.0 & 0.045 \\ 
RSGC\,2  & 279.838 & $-$6.029 & 42313 & 17872 & 13.0 & 16.0 & 0.021 & 12.0 & 15.0 & 0.024 & 11.0 & 14.0 & 0.028 \\ 
RSGC\,3  & 281.350 & $-$3.387 & 43557 & 20960 & 13.0 & 16.0 & 0.021 & 12.0 & 15.0 & 0.026 & 12.0 & 15.0 & 0.046 \\ 
F\,3 & 274.910 & $-$14.340 & 34837 & 15312 & 13.0 & 16.0 & 0.022 & 12.5 & 15.0 & 0.027 & 12.0 & 15.0 & 0.045 \\ 
F\,4 & 276.030 & $-$13.330 & 43445 & 19664 & 13.0 & 16.0 & 0.036 & 12.0 & 15.0 & 0.029 & 11.0 & 14.0 & 0.035 \\ \hline 
\end{tabular}
\end{table*}

\section{Data and Analysis Methods}\label{data}

\subsection{UKIDSS data}

In \citet{2013IJAA...03..161F} the main sequence of RSGC\,2 has been detected in
aperture photometry of UKIDSS GPS data between $K$\,=\,11\,mag and
$K$\,=\,14.5\,mag. All stars brighter than this are saturated. Tentatively, this
main sequence was also visible in 2MASS data starting from about
$K$\,=\,10\,mag. The red supergiants in this cluster are between 5$^{th}$ and
6$^{th}$ magnitude in the $K$-band \citep{2007ApJ...671..781D}. Thus, the
brightest main sequence members in this cluster are five or six magnitudes
fainter than the red supergiants. Given that several of the other known RSG
clusters are in a similar position on the sky and potentially have similar ages,
distances and reddening  (\citet{2006ApJ...643.1166F},
\citet{2007ApJ...671..781D}, \citet{2008ApJ...676.1016D},
\citet{2009A&A...498..109C}, \citet{2009AJ....137.4824A}) one can assume that
their main sequences should have analogue properties, i.e. should have similarly
bright members. However, for none of the other investigated RSG clusters or
candidates in \citet{2013IJAA...03..161F} has the main sequence been detected in
the GPS aperture photomety data. This was attributed to crowding in the
clusters, and thus low quality aperture photometry of potential main sequence
stars in the vicinity of the bright RSG cluster members. In other words, the GPS
aperture photometry catalogue is highly incomplete near the bright RSG cluster
stars. We hence selected RSGC\,1\,--\,5 and the cluster candidates F\,3 and F\,4
from \citet{2013IJAA...03..161F} for further investigations, i.e.
PSF-photometry, to reveal their main sequences.

We downloaded 10\arcmin\,$\times$\,10\arcmin\ cut-outs of the JHK images from
the UKIDSS GPS around the nominal position of each of these clusters via the
Widefield Camera Science
Archive\footnote{http://surveys.roe.ac.uk/wsa/index.html} webpage. We further
obtained a complete list of all source detections (independent of the
photometric quality) in each field. Typically there are about 30\,--\,45,000
detections in each of the fields. We then manually removed all detections which
were obvious image artefacts (detector cross talk, persistence, etc.) and added
all visible real stars missed by the source detection software to these
catalogues. In particular near the bright red supergiant cluster members a
number of real stars is missing. Several hundred detections have been
removed/added for each field by visually inspecting the $K$-band images and
manually deleting all obvious false detections and appending every object that
has the appearance of a star but was missing in the photometric catalogue. In
Table\,\ref{calibration} we list the number of objects in these input catalogues
for each cluster field.

\subsection{PSF Photometry}
 
We performed PSF fitting photometry in each image at all positions from the
above generated input catalogue. For this purpose, we used the standard routines
from DAOPHOT, as implemented in IRAF \citep{1987PASP...99..191S}. To define the
PSF for a given frame, about 10\,--\,15 isolated reference stars distributed
across the images were selected. Since the crowding was significant near the
field center, most of these reference stars are located in the outskirts of the
clusters. We did not notice any sign of a spatially varying PSF. The average PSF
of these reference stars is modelled in DAOPHOT with a combination of an
analytical function and an empirical image, which better represents the wings of
the PSF. This model PSF was then fit to all objects in the catalogue, to create
the final list of magnitudes. As measured by the $\chi^2$ and the PSF subtracted
images, the PSF fitting performs well in the UKIDSS GPS images. Only in one
case, the $K$-band image of the cluster candidate F\,3, is the PSF undersampled
and larger residuals are visible in the PSF subtracted images. However, the
quality of the photometry is not significantly degraded in this image (see
Table\,\ref{calibration}).

We merged the photometric catalogues for each of the JHK filters and used the
GPS aperture photometry of sources in each field to calibrate the instrumental
magnitudes. Only objects with {\tt pstar}\,$\ge$\,0.99965 (see
\citet{2008MNRAS.391..136L} for details on how this is defined) are used in the
calibration and a nominal shift of the magnitudes and no colour terms are
considered. The root mean square ({\it rms}) scatter of the calibrated
magnitudes is listed in Table\,\ref{calibration} together with the magnitude
range in which the calibration was performed. There are some images where we
detect a clear non-linearity for bright sources. This is expected, since the PSF
fitting should be reliable over a wider range of the nonlinear regime. Thus, the
colours and magnitudes for the bright stars could be systematically off. Note
that this only influences the very top of the potential cluster main sequences
and will have no influence on our analysis.

All point sources which did lack a detection by the PSF fitting routine in at
least one of the three NIR bands were removed from the final catalogues. We also
removed all stars which were fainter than the completeness limit (determined as
the peak in the magnitude distribution) in at least one filter. Finally all
objects that had a photometric uncertainty which differed by more than
2\,$\sigma$ from the average of stars with the same apparent magnitude were
removed. The number of stars in this output catalogue for each cluster is listed
in Table\,\ref{calibration}.

\subsection{Photometric Decontamination}

Using the calibrated JHK PSF-photometry we performed a photometric
decontamination of the stars in the cluster field to establish which stars are
the most likely cluster members. The method is based on the technique described
in \citet{2007MNRAS.377.1301B} and references therein. It uses the $J$-band
magnitudes and $J$$-$$H$ and $J$$-$$K$ colours to distinguish field stars from
cluster members based on their apparent magnitudes and colours. This particular
choice of colours provides the maximum variance among stellar cluster's
colour-magnitude sequences for open clusters of various ages
\citep{2004A&A...415..571B, 2007MNRAS.377.1301B}. We use a slight adaptation of
this method, outlined in detail in \citet{2010MNRAS.409.1281F}. For each cluster
we define as {\it cluster area} everything closer than the radius ($r$, as
specified in Table\,\ref{properties}) around the nominal cluster centre. Note
that the radii for the clusters are based either on literature estimates or are
chosen by us to include most of the confirmed or suspected RSG stars in each
object. The actual value of the radius will not influence our conclusions. As
{\it control field} we use all objects within the entire
10\arcmin\,$\times$\,10\arcmin\ field but further away than two cluster radii
from the centre of the cluster. See Fig.\,B1 in the Appendix for a
$K$-band and Glimpse 8\,$\mu$m image of RSGC\,1, 2, 3 with circles indicating
the cluster and control fields. We also show the $J$\,$-$\,$K$ vs $K$ colour
magnitude diagrams of the cluster and control fields for these clusters in
Fig.\,C1 in the Appendix. For each star ($i$) in the cluster
field we then determine the colour-colour-magnitude distance ($r_{ccm}$) to
every other star ($j$) in the cluster field as:

\begin{equation}\label{eq_rccm}
r_{ccm}=\sqrt{\frac{1}{2} \left( J_i - J_j \right)^2 + \left( JK_i - JK_j
\right)^2 + \left( JH_i - JH_j \right)^2}
\end{equation}

Where $JK = J-K$ and $JH = J-H$ are the above mentioned near infrared colours. 
The $r_{ccm}$ distance in which there are 20 stars in the cluster area is
denoted as $r^{20}_{ccm}$. It essentially defines the local density of stars in
the near infrared colour-colour-magnitude space. Note that the specific choice
of 20 stars does not influence any of our results and is a compromise between
the accuracy of the membership probabilities (see below) and the 'resolution' at
which we can determine the position of the main sequence. We then determine the
number of stars ($N^{con}_{ccm}$) at the same position and within the same
radius of the colour-colour-magnitude space but for the stars in the control
field. With this number, as well as the respective surface area of the control
field ($A_{con}$) and cluster area ($A_{cl}$), we can determine the
membership-likelihood index or cluster membership probability ($P^i_{cl}$) of
the star $i$ as:

\begin{equation}\label{eq_pcl}
P^i_{cl}=1.0-\frac{N^{con}_{ccm}}{20}\frac{A_{cl}}{A_{con}}.
\end{equation}

Note that these cluster membership probabilities are strictly speaking not real
probabilities (Buckner \& Froebrich 2013, subm.), since fluctuations of the
field star density in principle allow negative values for $P^i_{cl}$. If this
occurs in our analysis the $P^i_{cl}$ value is set to zero. However, the sum of
all $P^i_{cl}$ values equals the excess number of stars in the cluster area
compared to the control field. High values of $P^i_{cl}$ identify the stars in
the cluster field which are the most likely members and thus allow us to
establish the overall population of cluster stars statistically. Typically this
method identifies a few hundred stars in the fields of the clusters
RSGC\,1\,--\,3 which have a membership probability above 50\,\% (see later).

\begin{figure*}
\includegraphics[width=6.0cm, angle=-90]{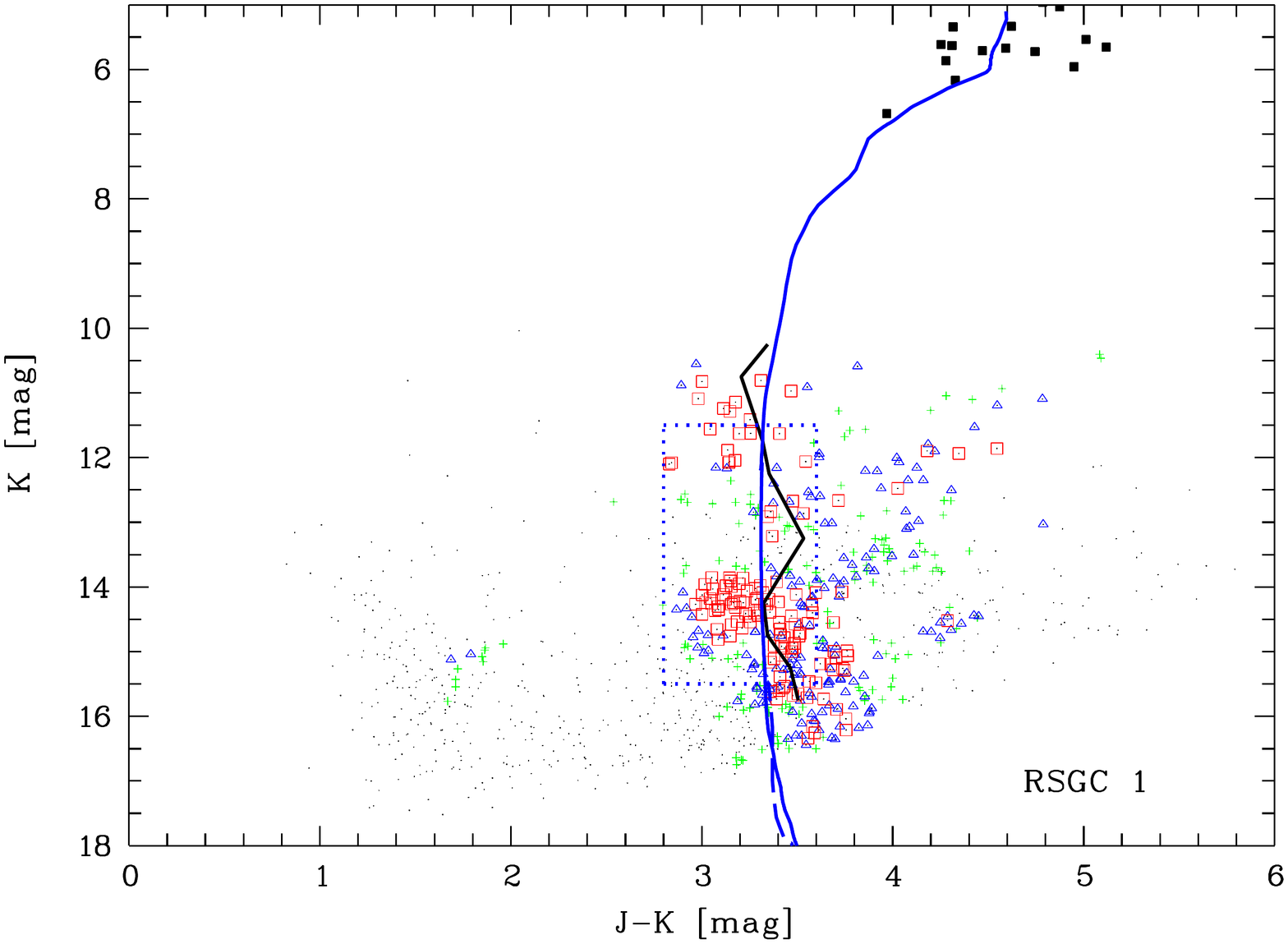} \hfill
\includegraphics[width=6.0cm, angle=-90]{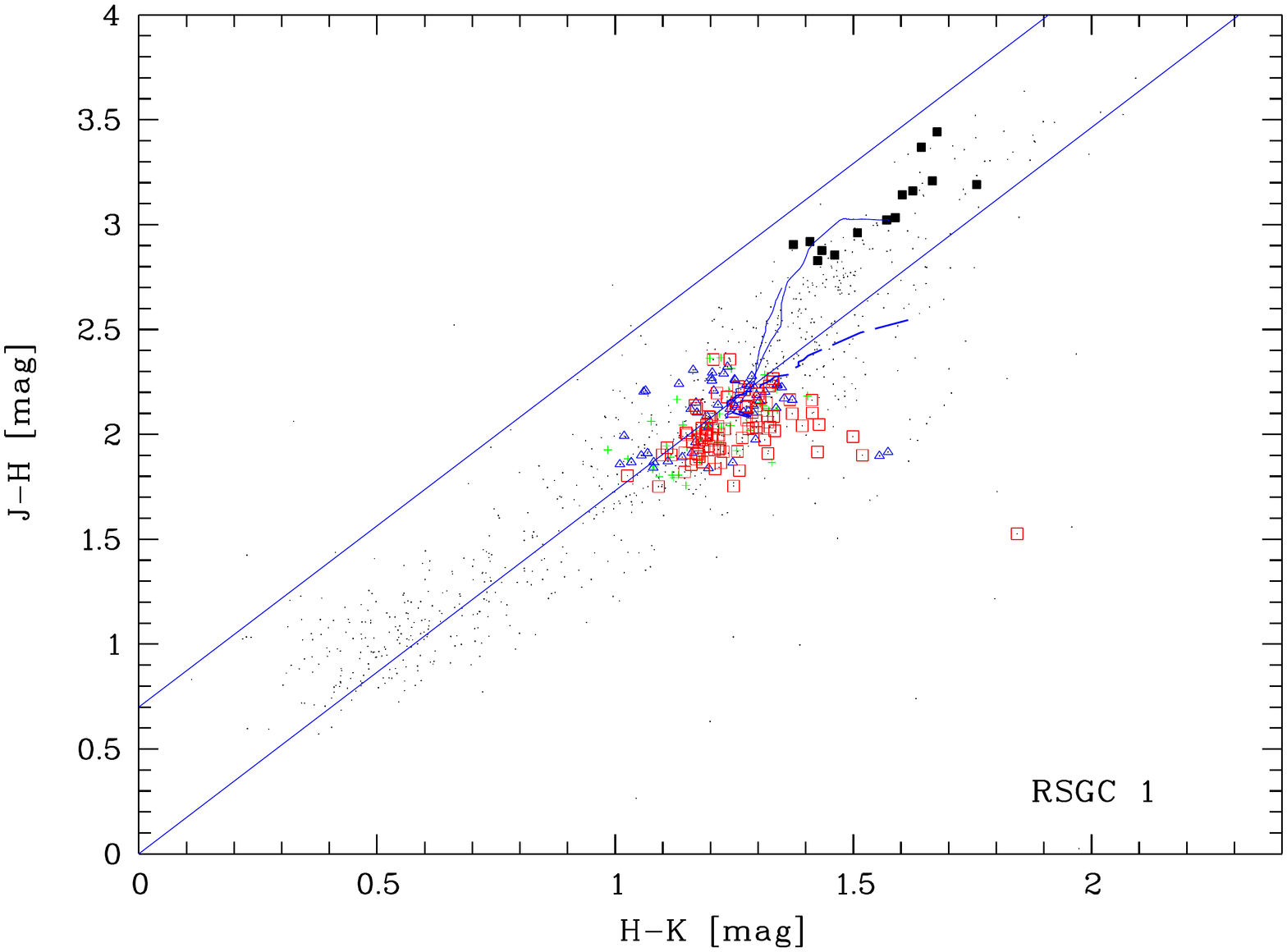} \\
\includegraphics[width=6.0cm, angle=-90]{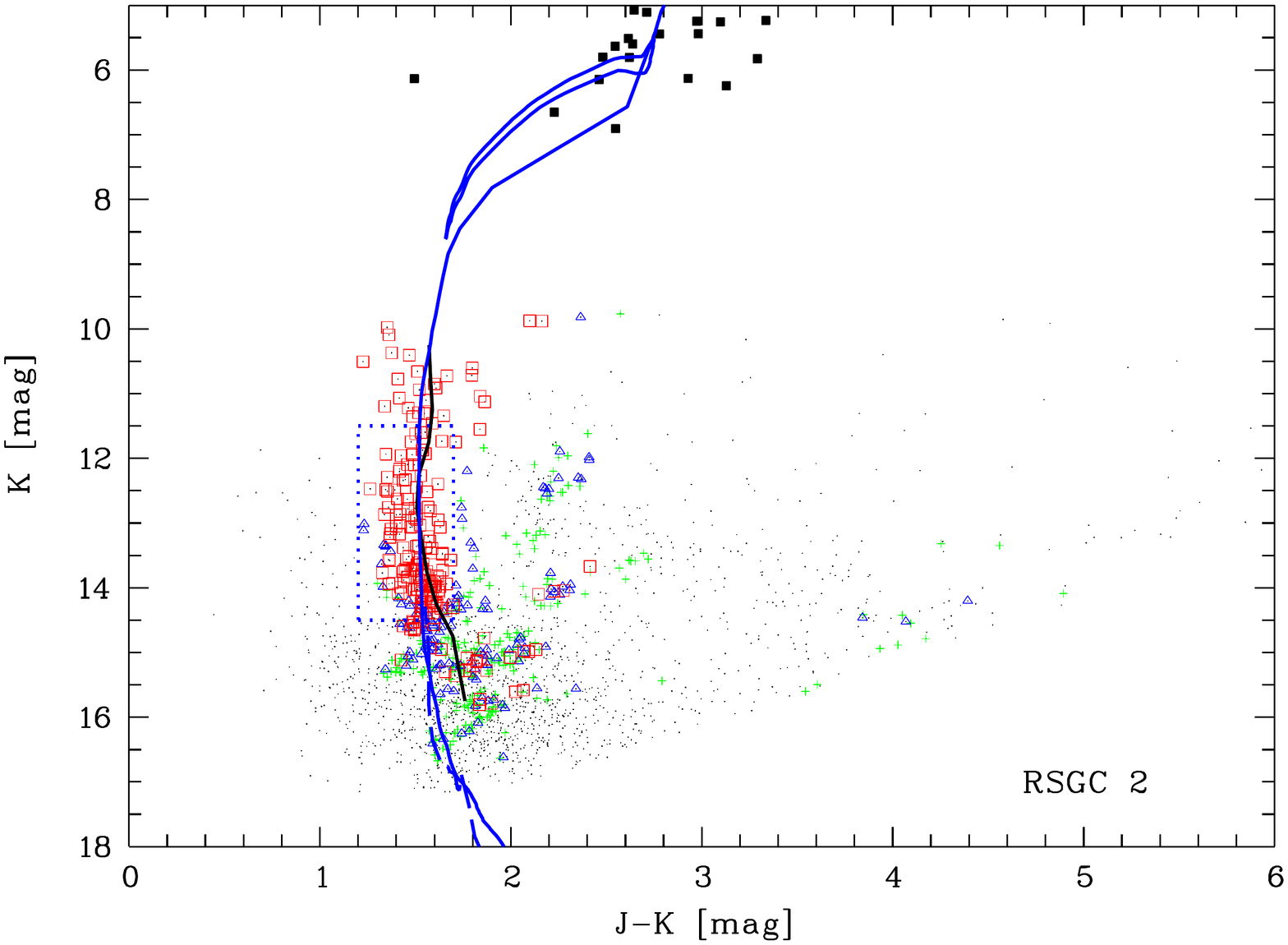} \hfill
\includegraphics[width=6.0cm, angle=-90]{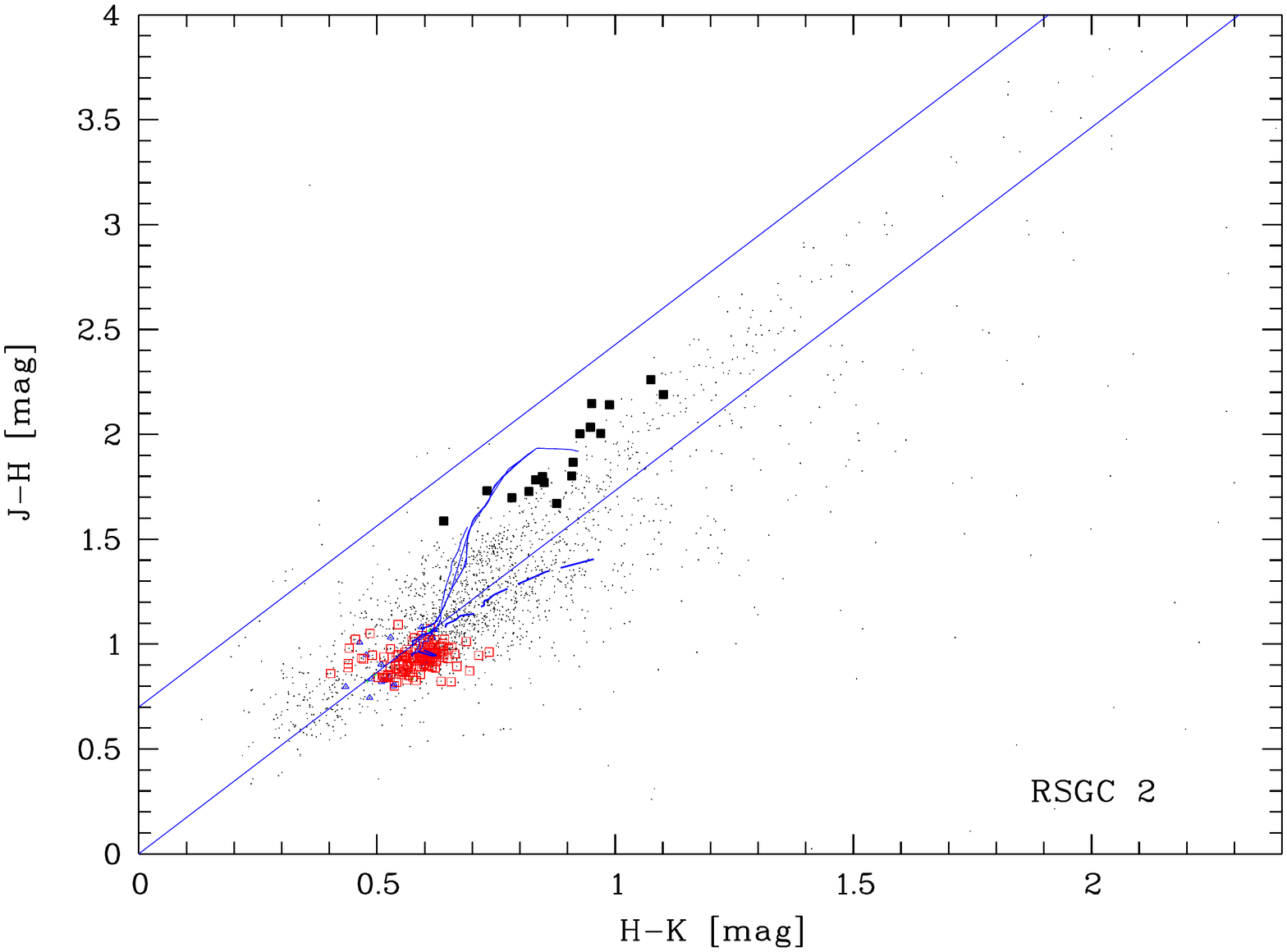} \\
\includegraphics[width=6.0cm, angle=-90]{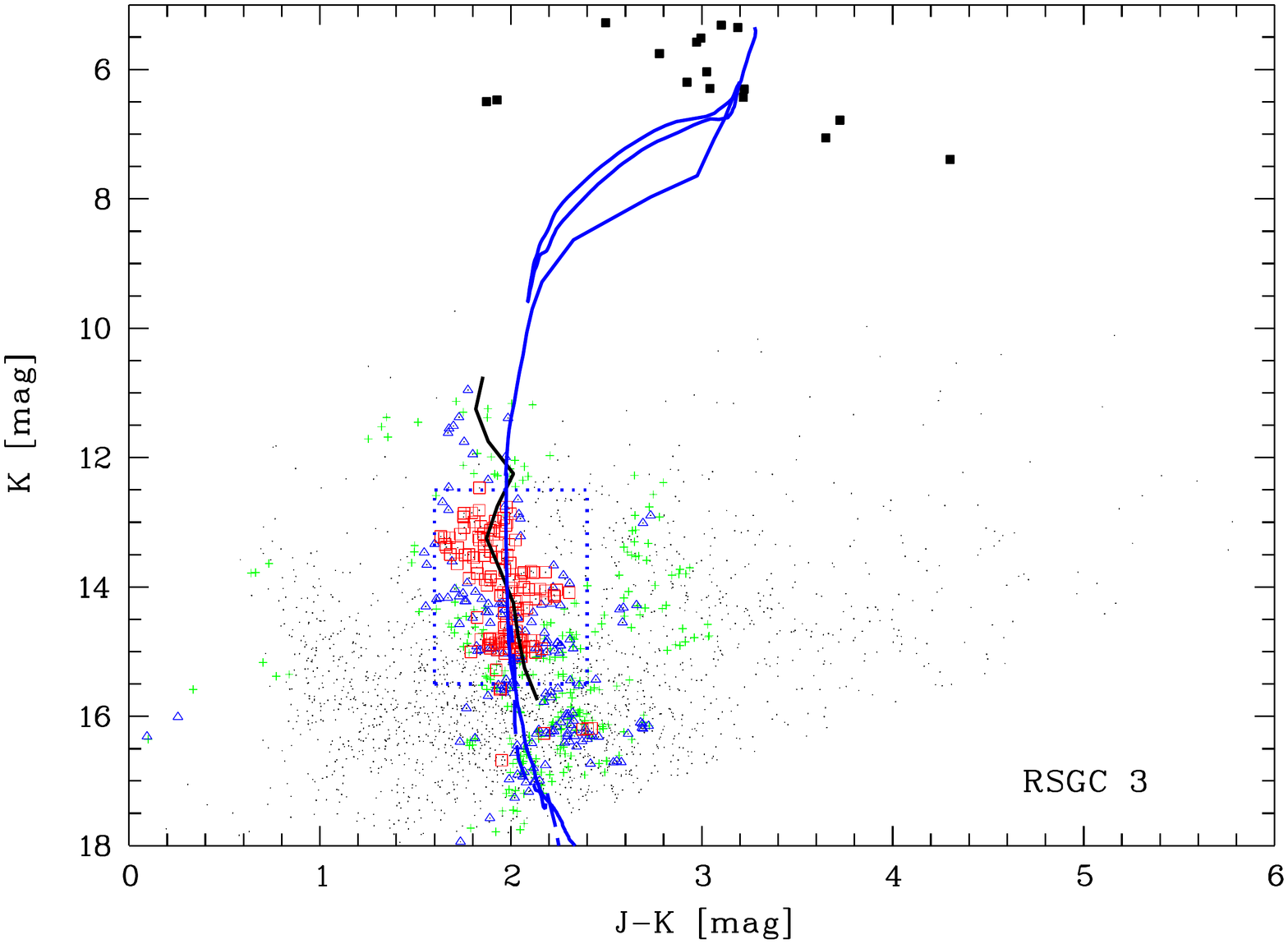} \hfill
\includegraphics[width=6.0cm, angle=-90]{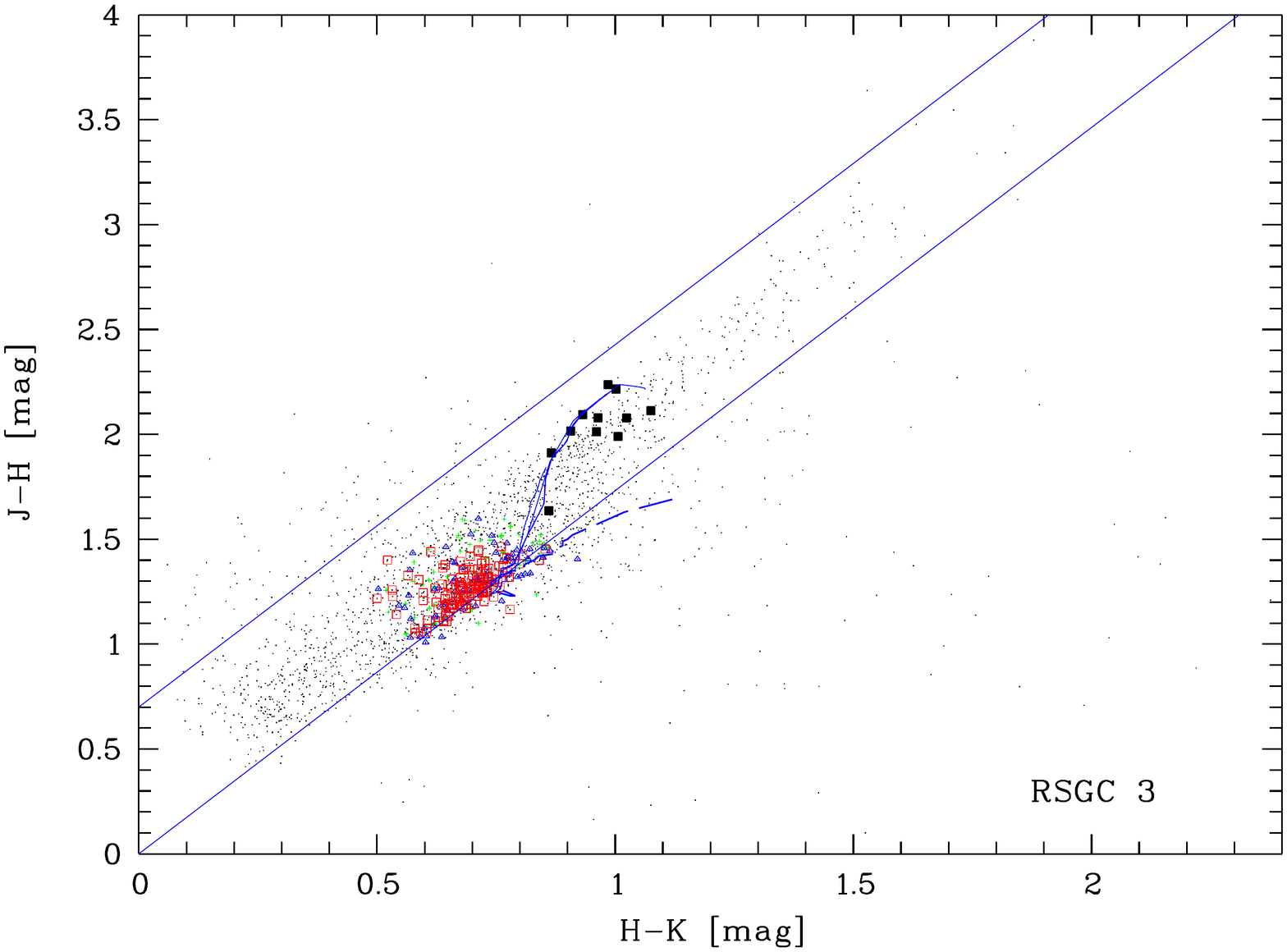} \\

\caption{\label{rsgcfig}  Colour-Magnitude (left) and Colour-Colour (right)
diagrams for RSGC\,1 (top), RSGC\,2 (middle) and RSGC\,3 (bottom). Objects
brighter than $K$\,=\,8\,mag are stars detected by 2MASS within 3\arcmin\ of the
cluster centre and are shown as black filled squares. The remaining stars are
from our PSF photometry. The colour coding indicates the photometric cluster
membership probability of the stars: $P^i_{cl} > 70\,\%$ red squares; $60\,\%
\leq P^i_{cl} < 70\,\% $ blue triangles; $50\,\% \leq P^i_{cl} < 60\,\% $ green
plus signs; $P^i_{cl} < 50\,\% $ black dots. In the CMDs, the black solid line
indicates the average position of the most likely cluster members, the blue
solid line an isochrone from \citet{2001A&A...366..538L} (using the ages listed
in Table\,\ref{properties}) and the blue dashed line an isochrone from
\citet{2000A&A...358..593S}. The dotted boxes indicate the region of stars
overplotted on the CCDs. The reddening band in the CCDs is based on the
extinction laws from \citet{2005ApJ...619..931I}.}

\end{figure*}

\begin{table*}
\centering

\caption{\label{properties}  Here we list the properties of the RSGC clusters
and the parameters of the isochrones overplotted on the CMDs and CCDs. We list
the radius $r$ out to which we include stars in the photometric decontamination,
the adopted distance and age of the cluster, as well as the $K$-band extinction
determined in the literature $A^{Lit}_K$ for the cluster RSGs and the extinction
$A^{MS}_K$ determined from our isochrone fit to the cluster main sequences.
$\Delta A^{MS}_K$ indicates the width of the main sequence, i.e. the
differential reddening. In $K^{3}_{MS}$ and $K^{10}_{MS}$ we list the apparent
$K$-band magnitude for which there are at least 3 (or 10) main sequence members
(weighted by their membership probability) per 0.5\,mag bin. The numbers in
brackets indicate the absolute magnitudes of these objects de-reddened with our
extinction. $N_{MS}^{>8\,M_\odot}$ is the number of potential massive main
sequence stars (weighted by their membership probability) brighter than
$M_K$\,=\,-0.9\,mag and $N_{RSG}$ the number of spectroscopically confirmed RSGs
in the cluster from the literature. See text for details.} 

\begin{tabular}{l|cccc|cccccc}
Name & r & d & age & $A^{Lit}_K$ & $A^{MS}_K$ & $\Delta$$A^{MS}_K$ & $K^{3}_{MS}$ & 
$K^{10}_{MS}$ & $N_{MS}^{>8\,M_\odot}$  & $N_{RSG}$\\ 
& [\,\,\arcmin\ ] & [kpc] &  [Myrs] & [mag] & [mag] & [mag] & [mag] & [mag]  & & \\ \hline 
\noalign{\smallskip}
RSGC\,1 & 1.5 & 6 & 10 & 2.74 & 2.3 &  0.75 & 10.5 (-5.9) & 12.5 (-3.7) & 210 & 14 \\ 
RSGC\,2 & 1.8 & 6 & 17 & 1.47 & 1.0 &  0.25 & 10.0 (-5.0) & 11.0 (-4.0) & 115 & 26 \\ 
RSGC\,3 & 1.8 & 6 & 20 & 1.50 & 1.4 &  0.25 & 11.5 (-3.6) & 13.0 (-2.1) & 115 & 16 \\ \hline 
\end{tabular}
\end{table*}


\section{Results}\label{results}

Here we discuss the results obtained for all the investigated clusters. Some
further details on the individual objects can be found in  Appendix\,A.

\subsection{General}

Theoretically one would expect the main sequence of these massive clusters to
appear as a vertical accumulation of high probability cluster members in a $K$
vs. $J$$-$$K$ colour-magnitude diagram (CMD), since most stars visible should be
of high mass and thus have similar intrinsic near infrared colours. Some scatter
in colour is expected due to differential reddening along the line of sight.
Furthermore, these high mass main sequence stars should be situated at the
bottom of the reddening band in a near infrared $H$$-$$K$ vs. $J$$-$$H$
colour-colour diagram (CCD) and not in the middle/top, which is usually occupied
by giant stars.

In the left column of panels in Fig.\,\ref{rsgcfig} we show as examples the
decontaminated $K$ vs. $J$$-$$K$ CMD of RSGC\,1, 2, 3. All stars fainter than
8$^{th}$ magnitude in $K$ are from our PSF photometry. All bright, potential
RSGs (large black squares) are taken from 2MASS since they are saturated in the
UKIDSS images. All small black dot symbols are stars in the cluster area with
less than 50\,\% cluster membership probability $P^i_{cl}$. Green plus signs
indicate stars with $50\,\% \leq P^i_{cl} < 60\,\% $, blue triangles stars with
$60\,\% \leq P^i_{cl} < 70\,\% $ and red squares $P^i_{cl} > 70\,\%$.

We determined a running weighted average of the $J$$-$$K$ colours of the most
likely cluster members along the detected main sequence. This is indicated by
the solid black line in the CMDs in Fig.\,\ref{rsgcfig}. As weighting factor for
each star we used the square of the membership probability $P^i_{cl}$. To
compare the cluster data with model isochrones, we utilise the Geneva isochrones
by \citet{2001A&A...366..538L} which are overplotted in each panel as a blue
solid line, using the parameters specified in Table\,\ref{properties}. We also
overplot as a dashed line the isochrones for low and intermediate mass stars
from \citet{2000A&A...358..593S}.

In the right column of panels in Fig.\,\ref{rsgcfig} we show the corresponding
$H$$-$$K$ vs. $J$$-$$H$ CCDs for the same three clusters. Symbols and colours
are identical in their meaning to the CMDs. However, we plot all stars in the
field as small black dots and only high probability cluster members ($P^i_{cl} >
50\,\%$) from the potential main sequences (as indicated by the dotted boxes in
the CMDs) are shown in large coloured symbols. These boxes exclude bright,
potentially saturated stars as well as faint, low signal to noise objects and
obvious background giants. The indicated reddening band for each cluster 
indicated, is based on the reddening law by \citet{2005ApJ...619..931I}.

From all the clusters investigated, we can detect a main sequence only in the
known objects RSGC\,1, 2 and 3 (see Fig.\,\ref{rsgcfig}). We also investigated
the fields around RSGC\,4 \citep{2010A&A...513A..74N} and RSGC\,5
\citep{2011A&A...528A..59N} but there are no apparent overdensities of stars, in
particular none that would indicate a main sequence (see Fig.\,D1 in the
Appendix). There are several possibilities that could explain this. i) These
clusters have less mass, i.e. fewer members, than the other objects and thus
they do not manifest themselves as overdensities in colour magnitude space. ii)
The clusters are much more extended spatially than our search area for potential
main sequence stars; they are more association like in appearance than cluster
like. Both points seem to contribute, since both clusters have fewer confirmed
members than the brighter RSGCs and they seem to be embedded in more extended
regions of massive young stars \citep{2010A&A...513A..74N, 2011A&A...528A..59N}.

For the new cluster candidates F\,3 and F\,4 from \citet{2013IJAA...03..161F}
only features that look like a tentative MS in the CMDs are found (see Fig.\,D2
in the Appendix).  For F\,3, a clump of stars at $K$\,=\,13\,mag and
$J$$-$$K$\,=\,2.4\,mag can be identified, while for F\,4 a more MS like feature
can be seen at $J$$-$$K$\,=\,2.6\,mag. Both of these features contain a few
hundred high probability members. However, when utilising the CCDs for both
cluster candidates, one can identify that these features are not caused by main
sequence stars. The high probability members clearly are not situated near the
bottom of the reddening band, indicating they are giants. Only in the case of
F\,3, there are some potential main sequence stars. Thus, both candidates are
most likely holes in the general extinction and not real clusters.

\begin{figure}
\centering
\includegraphics[width=6.2cm, angle=-90]{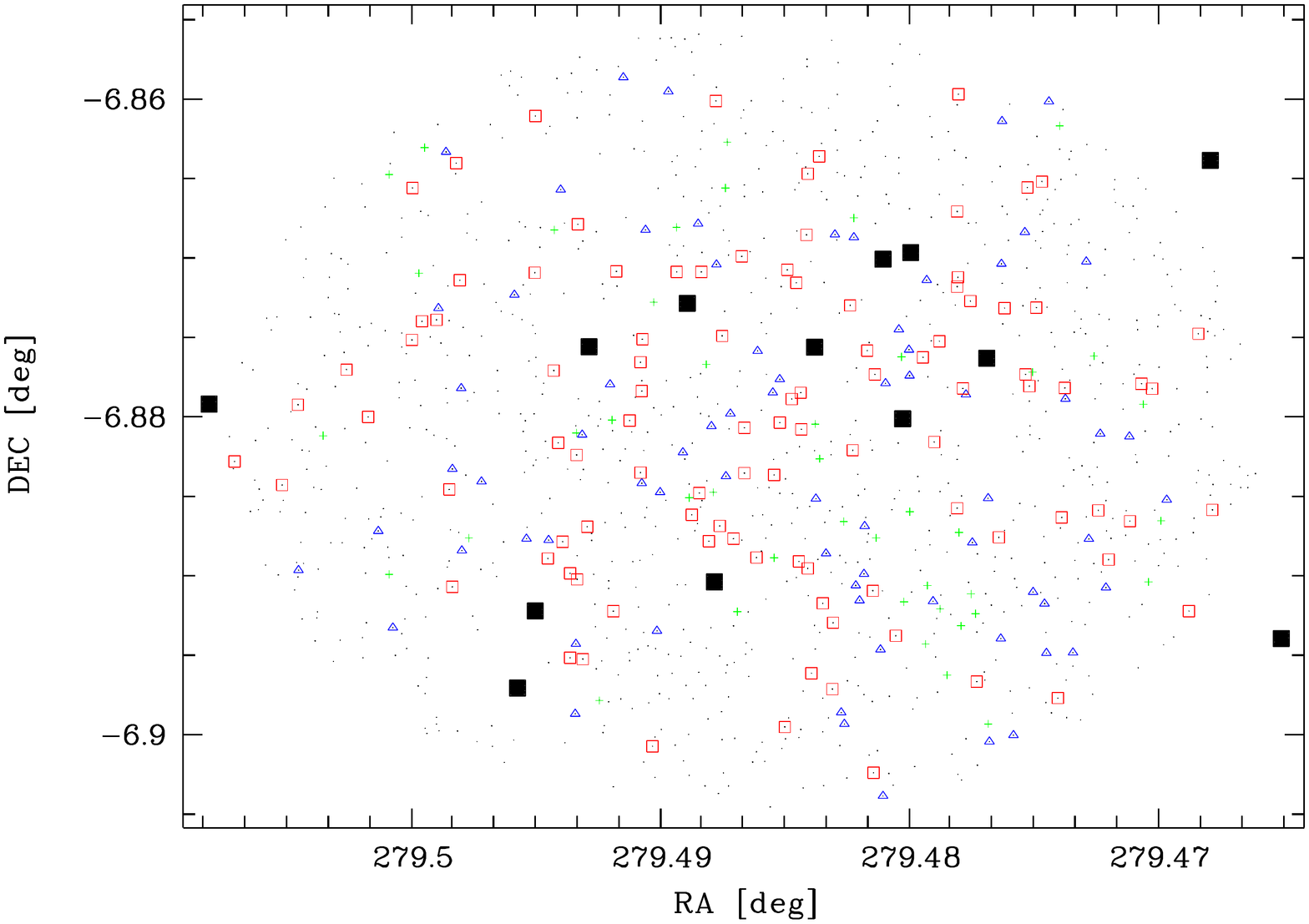} \hfill
\includegraphics[width=6.2cm, angle=-90]{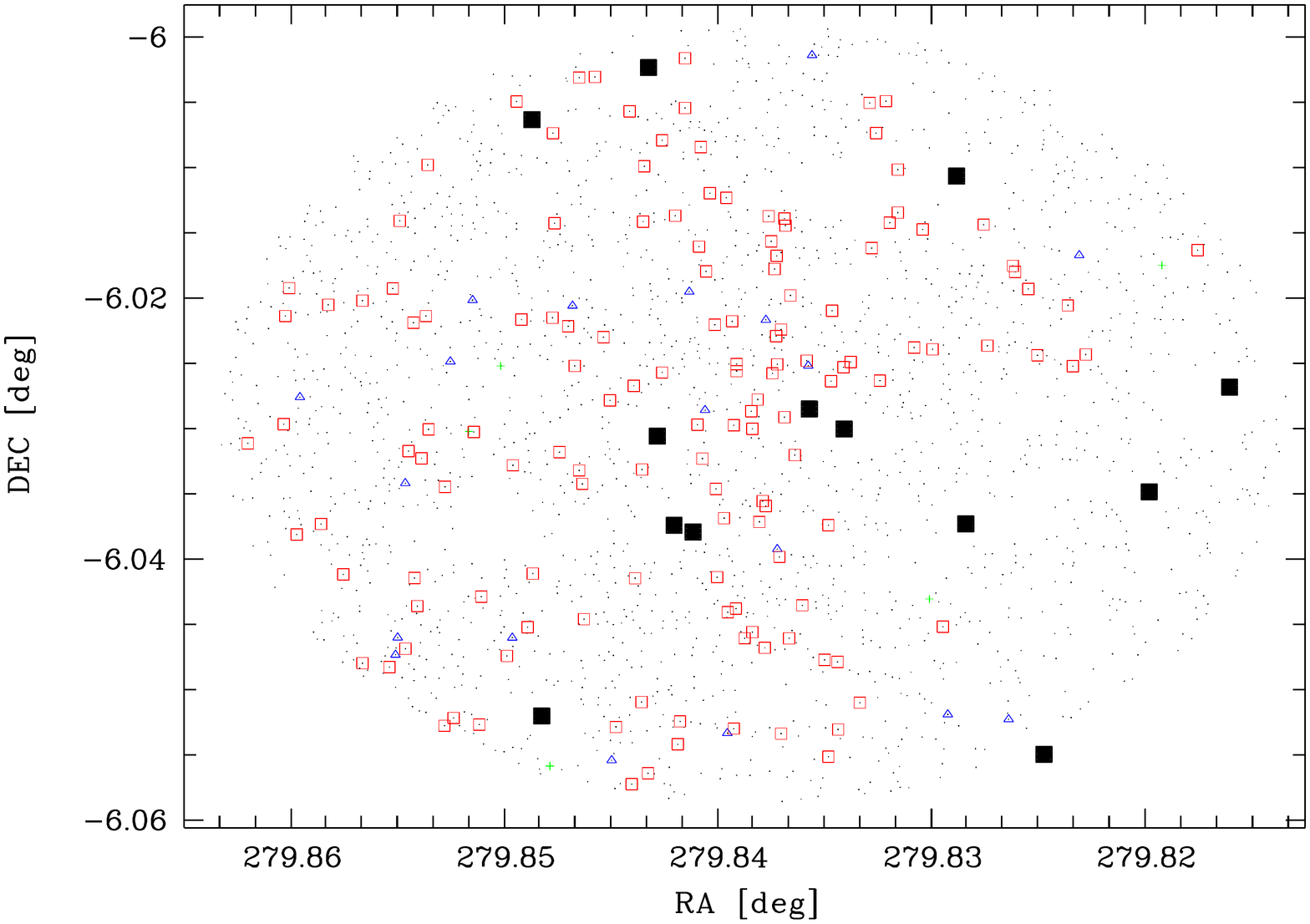} \hfill
\includegraphics[width=6.2cm, angle=-90]{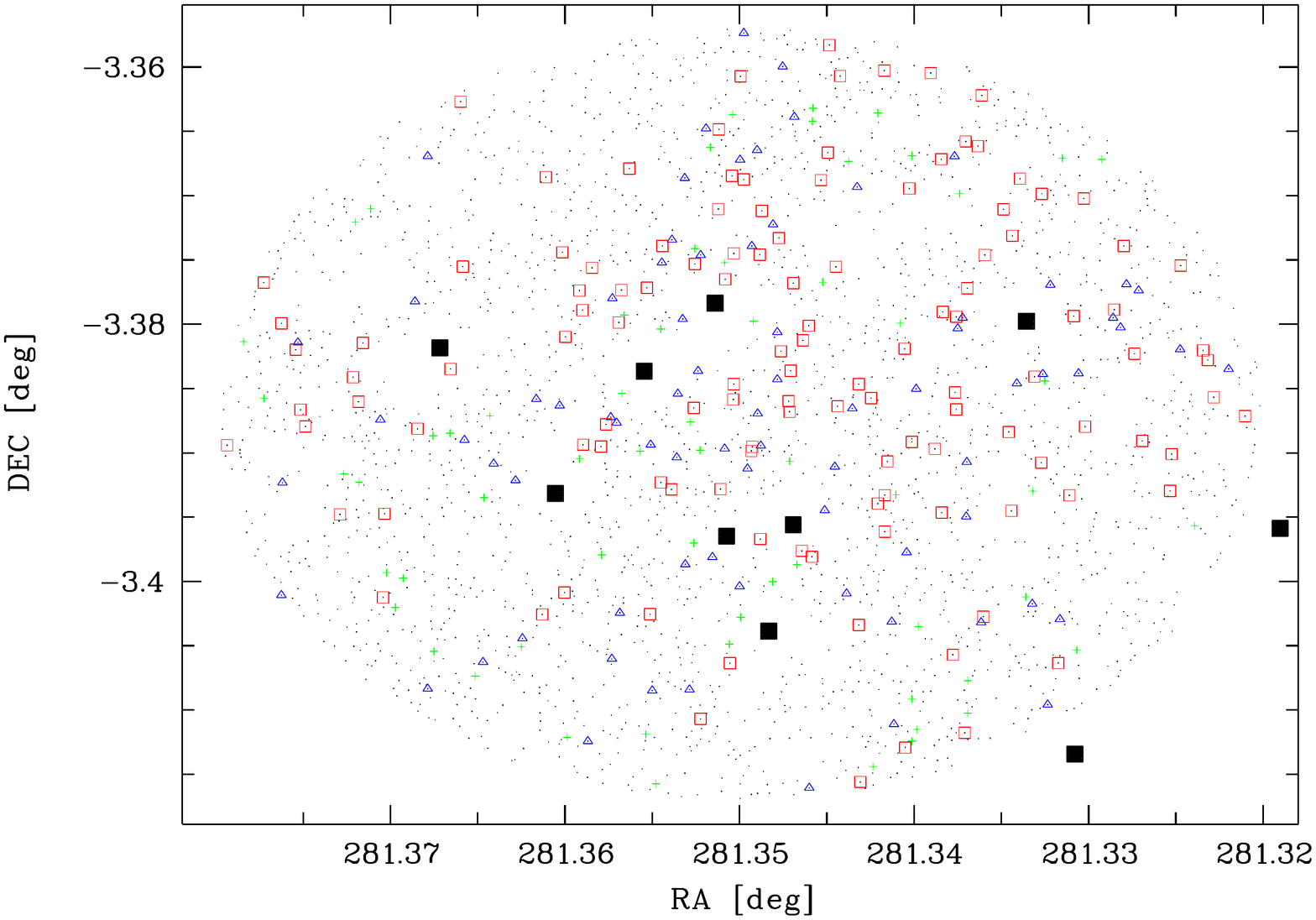} \\

\caption{\label{position_plot}  Positions of the stars in RSGC\,1 (top), RSGC\,2
(middle) and RSGC\,3 (bottom). Colour coding is the same as in
Fig.\,\ref{rsgcfig}. Only stars which are suspected MS stars (objects in the
blue dotted boxes in the CMDs in Fig.\,\ref{rsgcfig}) are plotted as coloured
symbols while the remaining stars in the field are represented by black dots and
confirmed RSGs are shown as black filled squares.} 

\end{figure} 

\subsection{Main Sequence Properties}

Here we will concentrate on determining the principle properties of the detected
main sequences for RSGC\,1, 2 and 3. The isochrones overplotted to the CMDs and
CCDs use age and distance estimates from the literature (see
Table\,\ref{properties} for these parameters). We adopt a distance of 6\,kpc for
all clusters, which seems an appropriate average of the published values
\citep{2006ApJ...643.1166F, 2008ApJ...676.1016D, 2007ApJ...671..781D,
2009A&A...498..109C, 2009AJ....137.4824A}. The ages are taken for each
individual cluster from the same references. We only vary the extinction in the
$K$-band to shift the isochrone onto the detected main sequence. Since the upper
end of the MS is almost vertical in the $K$ vs. $J$$-$$K$ CMDs, the actual
choise of age and distance will not influence the required extinction value.

All main sequences are 'vertical' in the CMDs for the top 2\,--\,4\,mag in the
$K$-band. Hence, these are clearly massive MS stars and we can try to estimate
the colour excess towards the cluster by shifting an isochrone until it fits the
MS. This will determine the extinction towards potential cluster members, i.e.
the column density of material along the line of sight that is not associated
with the cluster itself, and is independent of the actual age chosen for the
isochrone. Utilising an extinction law (we use \citet{2005ApJ...619..931I}), we
can convert this to a foreground extinction value. These are the values we used
to overplot the isochrones in Fig.\,\ref{rsgcfig} and which are listed as
$A_K^{MS}$ in Table\,\ref{properties}. In the colour-colour diagrams in the
right  hand side panels of Fig.\,\ref{rsgcfig} one can see that the reddening
direction of the background giants confirms the validity of this reddening law.
In essence the reddening law in these fields is in agreement with
\citet{2005ApJ...619..931I} or \citet{2009MNRAS.400..731S}, but using a less
steep dependence of extinction on wavelength such as in
\citet{1990ARA&A..28...37M} or \citet{1985ApJ...288..618R} can be ruled out from
the CCDs. Thus, please note that the use of an extinction law in agreement with
the CCDs will change the infered $K$-band extinction by only about 0.1\,mag.
However, larger differences are expected when a less steep extinction law is
applied. The cluster RSGC\,1 has the largest extinction of 2.4\,mag in the
$K$-band, while the other two clusters have about half this value of reddening,
i.e. $A_K$\,$\approx$\,1.2\,mag. Compared to the literature values (listed as
$A_K^{Lit}$ in Table\,\ref{properties}), our isochrone fit to the main sequence
systematically finds smaller extinction values. This could be caused by the fact
that: i) the literature extinction values are estimated using a different
extinction law (such as by \citet{2007ApJ...671..781D} for RSGC\,2); ii) the
observations are taken in different filters, e.g. 2MASS $K_S$ vs. UKIDSS $K$;
iii) the extinction values are determined from the spectral types of the red
supergiants in the cluster and not by the better understood main sequence stars.
Recently \citet{2013ApJ...767....3D} have shown that there can indeed by issues
with the RSG temperature scale.

We further estimate the amount of material associated with the cluster itself.
This can be done by measuring the width of the main sequence in $J$$-$$K$ and
convert this to a value of differential reddening. We list these values in the
$\Delta A_K$ column in Table\,\ref{properties}. As for the general interstellar
extinction, RSGC\,1 shows the highest amount of differential reddening with
about 0.75\,mag in the $K$-band. This is about three times as high as for the
other two clusters, but comparable to, or even smaller than for other young
embedded clusters (e.g. the Orion Nebula Cluster \citep{2011A&A...533A..38S}).
There are several possible explanations for the differences: i) This cluster is
younger and thus still more deeply embedded in its parental molecular cloud  --
however this is unlikely given the age of the cluster. ii) The differential
reddening is not caused by intrinsic dust, but by the variations in extinction
of the foreground material. The larger value for $A^{MS}_K$ for this cluster
could support this.

In order to investigate the number of potential main sequence stars in each
cluster we defined a colour range in $J$$-$$K$ for each cluster (as indicated in
the CMDs in Fig.\,\ref{rsgcfig}) that encloses all the potential MS stars. We
select all stars within this colour range and determine the $K$-band luminosity
function along the main sequence. These are shown as dotted blue lines in
Fig.\,E1 in the Appendix. If we only count the membership
probabilities $P^i_{cl}$ for each star, then we obtain a more realistic
luminosity function which is shown as solid red line in Fig.\,E1. The
smallest difference between the two luminosity functions is evident for RSGC\,2.
Thus, this is the cluster where the MS stands out most significantly from the
field stars. This is in agreement with the fact that this is the only cluster
where the main sequence has been detected previously
\citep{2013IJAA...03..161F}. The typical membership probability for the MS
cluster members is about 80\,\% for RSGC\,2, while it is of the order of 60\,\%
for the other two clusters.

We investigate the brightest MS stars in each cluster and define $K^{3}_{MS}$ as
the apparent $K$-band magnitude where there are at least three stars per
0.5\,mag bin along the MS. In other words we treat this as the top of the main
sequence. Similarly we define $K^{10}_{MS}$ and list both values in
Table\,\ref{properties}. The most populated cluster main sequence at bright
$K$-band magnitudes occurs in RSGC\,2. There, $K^{10}_{MS}$\,=\,11\,mag, while 
$K^{3}_{MS}$ is one magnitude brighter. RSGC\,3 has by far the fewest bright
cluster main sequence stars, or the MS starts only at fainter magnitudes. The
absolute magnitudes for $K^{3}_{MS}$ and $K^{10}_{MS}$ are determined from our
adopted distance and de-reddened with $A_K^{MS}$. They are listed in brackets in
Table\,\ref{properties} and show that RSGC\,1 has the brightest
($M_K$\,=\,-5.9\,mag) end of the MS while RSGC\,3 has intrinsically the faintest
end of the MS ($M_K$\,=\,-3.7\,mag). This is in good agreement with the ages for
the clusters determined in the literature, which range from 8\,--\,12\,Myrs for
RSGC\,1 \citep{2006ApJ...643.1166F,2008ApJ...676.1016D} to about 20\,Myrs for
RSGC\,3 \citep{2009AJ....137.4824A}. Note that according to the isochrones used
\citep{2001A&A...366..538L}, stars with a mass above 20\,M$_\odot$ (or O-type
stars) have $M_K$\,=\,-2.9\,mag or brighter on the MS. Thus, in particular
RSGC\,1 and RSGC\,2, could contain a significant number of massive MS or post-MS
objects, that can easily be verified spectroscopically. Note that the total
number of these cluster members is likely to be larger by about 50\,\%, since we
have only analysed the stars within one cluster radius. There are
spectroscopically confirmed cluster members in the region between one a two
cluster radii; typically only about 2/3$^{\rm rd}$ of the known members are
within one cluster radius.

The completeness limit determined as the peak of the $K$-band luminosity
function, for all clusters is between  $K$\,=\,15\,mag and 16\,mag. In all
cases, this is fainter than stars of about 8\,M$_\odot$ which have
$M_K$\,=\,-0.9\,mag \citep{2001A&A...366..538L} or an apparent magnitude of
$K$\,=\,13.0\,mag at our adopted distance and without considering extinction. We
thus can compare the total number of stars along the main sequence, brighter
than these stars. The numbers are weighted by the membership probabilities and
are listed as $N_{MS}^{>8\,M_\odot}$ in Table\,\ref{properties}. We find that
RSGC\,1  has about twice as many of these OB-type stars than RSGC\,2 and 3. 
Please note that we expect increased crowding in the cluster centres and thus
the estimated OB-type cluster member numbers should be treated as lower limits.
Since RSGC\,1 is the most compact of the three clusters, its numbers should be
most affected. The number of confirmed RSGs (see column $N_{RSG}$ in
Table\,\ref{properties}) in RSGC\,2 is much higher than in RSGC\,1, which might
be due to the lower age of the latter. Hence, out of the three clusters, RSGC\,1
seems to be the most massive object as estimated in \citet{2006ApJ...643.1166F}
and \citet{2008ApJ...676.1016D}. Based on the number of stars along the main
sequence and the age, RSGC\,2 and 3 might be less massive, partly (for RSGC\,3)
in agreement with the predictions from \citet{2009AJ....137.4824A} and at the
lower end of the mass range suggested in \citet{2009A&A...498..109C}. 

In Fig.\ref{position_plot} we show the spatial distribution of all stars along
the MS for the three clusters. Only stars within the cluster radius are shown,
since we have not determined membership probabilities outside this area. In all
three cases the most likely MS stars are concentrated towards the nominal centre
of the clusters. However, the distribution seems not be centrally condensed, but
rather filamentary, especially for RSCG\,2. If this is a real effect, or caused
by crowding in the cluster centre and 'missing' objects near the bright RSGs is
unclear. The spatial density of the main sequence cluster members (each star is
weighted by its membership probability) is between two and three times higher in
the cluster centre compared to the outer regions as defined by the cluster
radius. 

\section{Conclusions}\label{conclusions}

We have used PSF photometry on deep NIR JHK imaging data from the UKIDSS GPS to
investigate the fields of known and candidate red supergiant clusters. We
confirm the detection by \citet{2013IJAA...03..161F} of the upper main sequence
of the cluster RSGC\,2 and for the first time detect the main sequence for
RSGC\,1 and RSGC\,3. 

We use the age and distance estimates from the literature to overplot isochrones
on the NIR colour-magnitude diagrams for all clusters in order to establish the
reddening of the main sequences by utilising the reddening law from
\citet{2005ApJ...619..931I}. In all cases the infered $K$-band extinction values
for the main sequences are smaller than the quoted values in the literature,
which are determined for the red supergiant stars. We also infer the
differential reddening towards each cluster based on the width of the detected
main sequence. The youngest of the clusters (RSGC\,1) has the highest extinction
and differential reddening in accordance with its evolutionary status. It also
contains the most number of stars (about 200) with masses above 8\,$M_\odot$.

The spatial distribution of the candidate main sequence stars in all clusters
shows a concentration towards the nominal cluster centre. However, there is no
indication of a centrally condensed distribution, which could either be real or
caused by increased crowding and blending effects from the bright red supergiant
cluster members.

We also investigated fields near the clusters RSGC\,4 and RSGC\,5, as well as
the candidate RSG clusters F\,3 and F\,4 from \citet{2013IJAA...03..161F}. In
all cases no main sequence could be detected. In the case of the already known
clusters this could be caused by them having less members or being spatially
more extended, i.e. more association like. Our results indicate that the new
candidates can most likely be interpreted as holes in the background extinction.

\section*{Acknowledgements}

We would like to thank I.\,Negueruela and C.\,Gonzalez for fruitful discussions
during the earlier stages of the project. We further acknowledge the
constructive comments by the referee B. Davies which helped to improve the
paper. Part of this work was funded by the Science Foundation Ireland through
grant no. 10/RFP/AST2780.

 %
 %

\bibliographystyle{mn2e}
\bibliography{references}

\label{lastpage}

\end{document}